# Photo-induced valley currents in strained graphene.

T. L. Linnik

*Department of Theoretical Physics, V. E. Lashkaryov Institute of Semiconductor Physics, National Academy of Sciences of Ukraine, 03028 Kyiv, Ukraine*

**ABSTRACT**

The theoretical results are presented showing that strain-induced anisotropy of graphene spectrum gives rise to the valley currents under the illumination by normally incident light. The currents of the two graphene valleys are mutually compensated providing zero net electric current. The magnitude and direction of the valley currents are determined by the parameters of strain and light polarization. For not too high photon energy strain-induced valley current exceed that due to intrinsic warping of the graphene spectrum which suggests feasibility of strain-mediated valleytronics.

## 1. INTRODUCTION

Although importance of the valley structure of the carrier spectrum in crystals is recognized since the early age of solid-state physics, the idea to employ valley degree of freedom as an internal characteristic independent on electric charge and spin was formulated only recently [1]. The related theoretical concepts and first successful experiments suggest emerging of new direction called valleytronics. It assumes that in multi-valley crystals non-zero currents in individual valleys can be generated keeping zero total electric current. In experiments, the valley control is realized with the use of carrier photo-excitation in two-valley $MoS_2$ monolayer [2-4] and six-valley Si-based structure [5]. Moreover, valley Hall effect was observed recently for $MoS_2$ under the valley-selective optical excitation [6]. As graphene band structure has two inequivalent valleys, this material is a potential candidate for development of valleytronics. Although no experimental indication of valley currents in graphene is present so far, various approaches of their generation as well as valley filtering were proposed. The activity was started by the papers [7,8] where the specific valley-dependent edge states of graphene nanoribbons are proposed to be used for the valley filtering. Another approach explores valley Hall effect in graphene with lifted sublattice equivalence [9-11]. After that, a number of various approaches were suggested [12-26]. One of them, [23], employs warping of the graphene spectrum, which is essential at high carrier energies, above 1 eV. Such warping gives rise to valley currents under the optical excitation with light propagating normally to the graphene layer. It is important, that anisotropy of the graphene energy spectrum can be not only due to its intrinsic properties (warping), but also under application of the external strain [27-29]. In this paper we analyze the valley currents of illuminated strained graphene. It is known that application of strain to graphene conserves the Dirac form of its spectrum, but leads to essential anisotropy of the Fermi velocity. Theoretical and experimental analysis suggest that such a behavior is sustained for strain magnitude as high as 10% [30-33]. According to our results, application of strain gives rise to greater photo-induced valley currents for mid-infrared or softer illumination, in compare to that due to natural warping. In addition, it allows external control of the valley currents in graphene structures with tunable strain parameters.

It is worth to mention that other materials of graphene family can also posses strain-induced spectrum asymmetry (see, for example [34] where the spectrum of strained bi-graphene was addressed), which suggests their perspectives for strain-controlled valleytronics.

## 2. SPECTRUM OF GRAPHENE UNDER UNIFORM STRAIN

The honeycomb crystal lattice of unstrained graphene and corresponding first Brillouin zone are shown in figures 1 (a) and (b), respectively. The Brillouin zone extrema are at two inequivalent corners of the hexagon, K and K'. The effect of the uniform strain on the energy spectrum of graphene was initially explored within the tight-binding approach and first-principles calculations [27,35-42]. The main results were that the opening of a gap in the energy spectrum requires very high values of strain, of the order of 20%. This means that the energy spectrum remains gapless and cone-like for moderate uniform strains. However, the Dirac points in strained graphene no longer coincide with the edges of the Brillouin zone, K and K'. Moreover, the strong strain-induced anisotropy of Fermi velocity appears [27-29,35,37]. On the other hand, the properties of intrinsic graphene [43,44] and graphene subject to various fields, [28,45-51], can be addressed based on the symmetry considerations. In particular, the **kp** Hamiltonian of strained graphene can be developed [28,47,48]. It results in the Dirac-like electron and hole spectra $E_{\mathbf{k}}^{(c,v)}$ with anisotropic electron and hole Fermi velocities:

$$E_{\mathbf{k}}^{(c,v)} = \hbar(\pm v_0(\varphi) + \delta v(\varphi))k . \tag{1}$$

Here $k$ and $\varphi$ are the absolute value and polar angle of momentum **k**, + and − signs correspond to the conduction and valence bands, and we dropped inessential for our problem strain-related momentum and energy shift of the Dirac point. In terms of the uniaxial, $\varepsilon_\Delta = \varepsilon_{xx} - \varepsilon_{yy}$, hydrostatic, $\bar{\varepsilon} = \varepsilon_{xx} + \varepsilon_{yy}$, and shear, $\varepsilon_{xy}$, components of strain we have

$$v_0(\varphi) = v_F (1 + \bar{\varepsilon}\tilde{d}_2/2 + \varepsilon_\Delta \tilde{g}_2 \cos 2\varphi/2 + \varepsilon_{xy}\tilde{g}_2 \sin 2\varphi), \qquad (2)$$

$$\delta v(\varphi) = 2v_F a_2 (\varepsilon_\Delta \cos\varphi - 2\varepsilon_{xy} \sin\varphi).$$

Here Fermi velocity in unstrained graphene $v_F = 10^6\ m/s$. The coefficients $a_2 \approx 0.2$, $\tilde{d}_2 \approx -1.25$ and $\tilde{g}_2 \approx -2.14$ are responsible for the anisotropy of the Fermi velocity and were determined in [28] by the comparison with the first-principles calculations. As we see, the energy spectrum of strained graphene is essentially anisotropic and strain breaks not only the equivalence of $\mathbf{k}$ and $-\mathbf{k}$ directions but also the symmetry of electron and hole spectra as it is shown in Fig. 2. The corresponding solution for the wave function is:

$$\Psi_k^{(c,v)}(\varphi) = \frac{1}{\sqrt{2}}\begin{pmatrix} 1 \\ \pm C(\varphi) \end{pmatrix} e^{i\mathbf{k}\mathbf{r}}, \qquad C(\varphi) = e^{i\varphi}\frac{1 + \tilde{d}_2\bar{\varepsilon}/2 + \tilde{g}_2(\varepsilon_\Delta/2 + i\varepsilon_{xy})e^{-i2\varphi}}{\sqrt{1 + \tilde{d}_2\bar{\varepsilon} + \tilde{g}_2\varepsilon_\Delta \cos 2\varphi + 2\tilde{g}_2\varepsilon_{xy} \sin 2\varphi}}, \qquad (3)$$

where, as in (1), $+$ and $-$ signs mark the carrier bands. The provided spectra and wave functions are for K valley. For K' valley, the expressions for the spectrum and wave functions can be obtained by substitution $x \to -x$ for momentum and strain components, or, explicitly, $\varphi \to \pi - \varphi$ and $\varepsilon_{xy} \to -\varepsilon_{xy}$.

## 3. LIGHT-INDUCED VALLEY CURRENTS

### 3.1. Photo-generation valley currents.

Before proceeding to rigorous analysis of the valley currents, let us discuss qualitatively its physical origin. In general, valley current can appear due to anisotropy of the carrier group velocity and photon-induced transition probabilities. In Fig.2 we plot the spectra of unstrained (red line) and strained (black line) graphene along $k_y$ direction for the case of pure shear strain, $\varepsilon_{xy}$. The vertical arrows show light-induced electron transitions from the valence to the conduction bands in K and K' valleys. The arrows correspond to the $y$-components of the photo-generated electron and hole group velocities. As we see, in unstrained graphene the resulting current in each valley is exactly zero. In the presence of strain this is still true if the probabilities of transitions at positive and negative $k_y$ are equal. However, as we will see below, this is not the case (in the figure this is illustrated by the different vertical arrow thicknesses). As a result, each valley possesses non-zero current. These currents in the K and K' valleys are anti-parallel, and there is no total current in the system.

The quantitative consideration of both effects can be done with the use of the steady state quasiclassical kinetic equation:

$$J^{(i)}\{f\} + J_R^{(i)}\{f\} + G^{(i)}\{f\} = 0, \tag{4}$$

where $f$ is the carrier distribution function, $J^{(i)}$ is scattering integral, $J_R^{(i)}$ and $G^{(i)}$ are recombination and interband photo-generation rates, and index $i$ marks the valley. We concentrate on the case of moderate temperatures and excitation photon energy below the doubled inter-valley (about 157meV, zone-edge transverse phonon mode) energy [52]. In this case we can neglect by inter-valley scattering, and the kinetic equations for each valley are decoupled. In the following we analyze kinetic equation for K valley and drop the valley index for all values discussing the results for K' valley at the end of the section. We assume also that actual carrier energies are less than that of optical phonon (about 200meV). Thus, we can also neglect by the optical phonon scattering and take $J\{f\} = J_{LA}\{f\} + J_{ee}\{f\} + J_{im}\{f\}$, considering scattering due to the longitudinal acoustic phonons (LA), impurity scattering (im), and electron-electron scattering (*ee*). Below we consider both intrinsic and doped graphene. However, we always assume that optical excitation generates carriers away from the Fermi energy level and we deal with fully populated initial carrier state and empty finite state. Formally, this means that $G$ does not depend on the distribution function. In the presence of strain both wave functions and light-electron interaction Hamiltonian [23] are modified, and we have

$$G = C_{eff} \sum_{\mathbf{k'}} \left| \langle \Psi_{\mathbf{k'}}^{(c)} | (\boldsymbol{\sigma} \cdot \mathbf{u} + \delta H_{u\varepsilon}) | \Psi_{\mathbf{k}}^{(v)} \rangle \right|^2 \delta(E_{\mathbf{k'}}^{(c)} - E_{\mathbf{k}}^{(v)} - \hbar\omega), \qquad C_{eff} = 16\pi^2 v_F^2 \alpha I_o t^2 / \omega^2, \quad (5)$$

were $\sigma_i (i = x, y)$ are the Pauli matrices, $\alpha = e^2/4\pi\varepsilon_0\hbar c \approx 1/137$ is dimensionless fine structure constant; $I_0$, $\mathbf{u}$ and $\omega$ are the intensity, polarization and the frequency of the incident light, respectively. We introduce also here the electric field amplitude transmission coefficient $t = 2/(n+1)$ assuming the graphene sheet is placed at the substrate with the refractive index $n$. $\Psi_{\mathbf{k}}^{(c,v)}$ are the wave functions of the conduction and valence bands which are determined by Eq. (3). The strain-induced contribution to the light-electron interaction, $\delta H_{u\varepsilon}$, is analogous to the $H_{k\varepsilon}$ terms in the Hamiltonian of strained graphene [28] and is determined by the same constants:

$$\begin{aligned} \delta H_{u\varepsilon} &= a_2((u_x + iu_y)(\varepsilon_\Delta + 2i\varepsilon_{xy}) + (u_x - iu_y)(\varepsilon_\Delta - 2i\varepsilon_{xy}))I + \\ &+ \tilde{d}_2 \bar{\varepsilon}((u_x - iu_y)\sigma_+ + (u_x + iu_y)\sigma_-)/2 + \\ &+ \tilde{g}_2((u_x + iu_y)(\varepsilon_\Delta - 2i\varepsilon_{xy})\sigma_+ + (u_x - iu_y)(\varepsilon_\Delta + 2i\varepsilon_{xy})\sigma_-)/2. \end{aligned} \qquad (6)$$

In linear in strain approximation $\delta H_{u\varepsilon}$ makes no contribution to the valley currents. Therefore, to avoid dealing with cumbersome expressions, we omit below the corresponding terms.

To solve Eq.(4) we use the standard approach [53], introducing as independent variables of the distribution function energy, $E$, and $\varphi$, and expanding $f$ into the Fourier series:

$$f(E,\varphi) = f_0(E) + \sum_{n=1}^{\infty} \left( f_n^{(c)}(E)\cos(n\varphi) + f_n^{(s)}(E)\sin(n\varphi) \right). \quad (7)$$

In these variables the generation term is

$$G = \text{sgn}(E)\frac{C_{\text{eff}}}{4}\left[1 + (u_y^2 - u_x^2)\cos(2\varphi) - 2u_x u_y \sin(2\varphi)\right]\delta\left(|E|(1 - \text{sgn}(E)\delta v(\varphi)/v_F) - \hbar\omega/2\right). \quad (8)$$

In the following, we assume that elastic scattering on various kinds of defects is the most efficient one. For the elastic scattering integral calculated assuming no strain effect on the carrier scattering probabilities, $J_{im}^{(0)}$, we have $J_{im}^{(0)}\{f\} = -\sum_{n=1}^{\infty}\left(f_n^{(c)}(E)\cos(n\varphi) + f_n^{(s)}(E)\sin(n\varphi)\right)/\tau_n(E)$, where $\tau_n$ are determined by the elastic scattering probabilities [53]. Since $J_{im}^{(0)}\{f\}$ contains no zero harmonic, $f_0$ is controlled by the other, less efficient, scattering mechanisms and we may assume that $f_0 \gg f_{n \neq 0}$.

Then, we introduce expansions $G = \sum_{n=0}^{\infty}\left(G_n^{(c)}(E)\cos(n\varphi) + G_n^{(s)}(E)\sin(n\varphi)\right)$,

$J_\nu\{f_0\} = \sum_{n=0}^{\infty}\left(J_n^{(\nu,c)}(E)\cos(n\varphi) + J_n^{(\nu,s)}(E)\sin(n\varphi)\right)$, where $\nu$ marks the scattering mechanisms, including recombination. Note, that for elastic scattering $J_{im}\{f_0\} = 0$. As a result, for $f_{n \neq 0}$ we have an approximate equation

$$f_n^{(c,s)}(E) = \tau_n(E)\left(G_n^{(c,s)}(E) + \sum_v J_n^{(v,(c,s))}(E)\right). \tag{9}$$

Note, that $J_n^{(v,(c,s))}$ for $n \neq 0$ appear only due to strain-induced anisotropy. In the following, we disregard this effect for the phonon, impurity and electron-electron scattering. In general, it leads to the corresponding contributions to $f_n^{(c,s)}$ and, consequently, to the valley current. Those contributions are difficult to analyze quantitatively since they depend, in particular, on the microscopic details of the phonon scattering and peculiarities of many-electron effects under the electron-electron scattering. Thus, we assume

$$f_n^{(c,s)}(E) = \tau_n(E)\left(G_n^{(c,s)}(E) + J_n^{(R,(c,s))}(E)\right), \tag{10}$$

and the corresponding valley current could be treated as a lower estimate.

Using the $E, \varphi$ variables, the expression for the valley current is

$$j_i = \frac{e}{2\pi^2 \hbar^2} \int dE d\varphi \frac{E \operatorname{sgn}(E) v_i^{(g)} f(E,\varphi)}{v_E^2(\varphi)}, \quad i = x, y, \tag{11}$$

where $\mathbf{v}^{(g)} = \hbar^{-1} \nabla_k E$ is the carrier group velocity and $v_E(\varphi) = \operatorname{sgn}(E) v_0(\varphi) + \delta v(\varphi)$. According to (10), we can split the valley current into generation and recombination contributions, $j_i^{(G)}$ and $j_i^{(R)}$ calculated by (11) for the corresponding contributions of the distribution functions, $\tau_n(E) G_n^{(c,s)}(E)$ and $\tau_n(E) J_n^{(R,(c,s))}(E)$. Since in our model $G$ does not depend on $f_0(E)$, it is straightforward to obtain explicit expression for $j_i^{(G)}$ which are valid for both intrinsic and doped graphene. This is not true for $j_i^{(R)}$. We postpone the related analysis of $f_0(E)$ and $j_i^{(R)}$ till the next section concentrating here on calculation of $j_i^{(G)}$.

The nonzero contribution to $j_i^{(G)}$ is provided by the zero harmonic of the factor $f(E,\varphi)v_i^{(g)}(E,\varphi)/v_E^2(\varphi)$. Restricting ourselves by the first-order contribution with respect to the magnitude of strain, we conclude that there are two inputs to the valley current. The first one is due to $f_1^{(c,s)}$, which appears under the expansion of the $\delta$-function in (8). The second one is first harmonic of $v_E^{-2}$. Both these contributions are stipulated by the strain-induced asymmetry of the electron and hole spectra, manifested in the anisotropy of the transition probability and effective density of states under the photon-induced transitions. As a result, we obtain for the electron part of $j_i^{(G)}$

$$j_x^{(G)} = \frac{\pi e a_2 v_F \alpha I_o t^2}{E_\omega^2}\left\{\left[\varepsilon_\Delta(u_y^2 - u_x^2) + 2\varepsilon_\Delta + 4\varepsilon_{xy}u_xu_y\right]\frac{d}{dE}\left\{E^2\tau_1(E)\right\}\bigg|_{E=E_\omega} + \right.$$
$$\left. + 2E_\omega\tau_2(E_\omega)\left[\varepsilon_\Delta\left(u_y^2 - u_x^2\right) + 4\varepsilon_{xy}u_xu_y\right]\right\},$$
$$j_y^{(G)} = \frac{2\pi e a_2 v_F \alpha I_o t^2}{E_\omega^2}\left\{\left[\varepsilon_{xy}(u_y^2 - u_x^2) - 2\varepsilon_{xy} - \varepsilon_\Delta u_xu_y\right]\frac{d}{dE}\left\{E^2\tau_1(E)\right\}\bigg|_{E=E_\omega} + \right.$$
$$\left. + 2E_\omega\tau_2(E_\omega)\left[\varepsilon_{xy}\left(u_y^2 - u_x^2\right) - \varepsilon_\Delta u_xu_y\right]\right\},$$
(12)

where $E_\omega = \hbar\omega/2$. For the hole part, we have analogous expressions but with minus sign and substitution $E_\omega \to -E_\omega$. To obtain the expression for photocurrent in K' valley we must change the sign of all $x$-components for all vectors ($\mathbf{j}^{(G)}$ and $\mathbf{u}$) and $\varepsilon_{xy}$. This means that the partial valley currents of two valleys have opposite signs being equal in magnitude which results in a zero net electric current in accordance with the symmetry arguments.

Let us proceed with some quantitative estimates. First of all, it is worth to compare the strain-induced valley current and that due to natural warping of the graphene spectrum [23]. To be specific, we assume elastic scattering by the unscreened Coulomb impurities in intrinsic graphene where $\tau_1 \sim E$ and $\tau_2 = 3\tau_1$ [23]. Taking parameters of warping from [23], for light polarized along $y$ direction and $\varepsilon_{xy} = 0$ the ratio of the warping, $j_x^{(w)}$, and strain-induced valley currents is

$$\frac{j_x^{(w)}}{j_x^{(G)}} = \frac{1}{96 a_2 \varepsilon_\Delta} \frac{\hbar \omega}{E^*}, \tag{13}$$

where $E^* \approx 17.5 \text{eV}$ denotes the energy where characteristic warping and Dirac contributions to the spectrum are comparable [23]. Thus, for realistic strain $\varepsilon_\Delta = 1\%$ strain-induced valley current exceeds the warping one for $\hbar \omega < 1\, eV$. This means that for long-wavelength radiation, starting from mid-infrared band, it is feasible to deal with strain-controlled valley current. Taking $\hbar \omega = 0.4\, eV$ and $\tau_{1\omega} = 10^{-14} s$ [54], for the provided above strain, light intensity $I_0 = 10^2\, \text{W/m}^2$ and the substrate refractive index $n = 2.6$, corresponding to SiC, we estimate $j_x^{(G)} = 2 \cdot 10^{-3}\, \text{pA}/\mu m$.

## 3.2. Recombination-induced valley currents.

In analogy to the considered above photo-generated valley currents the strain-induced anisotropy of the energy spectrum leads also to the appearance of the valley currents due to the inverse recombination processes. In general, a number of recombination processes are possible in graphene, including radiative, phonon-assisted, and Auger process [55]. For the considered excitation energy optical-phonon-assisted recombination is suppressed, while the Auger recombination is inefficient (see [56]). Thus, we concentrate on the former mechanism where spontaneous and thermal radiation-induced interband transitions take place. To estimate this effect we use the collision integral for the thermal radiation interband transitions given in Ref.

[57]. So, for positive energies corresponding to the conduction band, an explicit expression for $J_R$ is

$$J_R\{f(E,\varphi)\} = v_E^{(R)}\left[N_{ph}(E,\varphi)(1-f(E,\varphi))f(E',\varphi) - (N_{ph}(E,\varphi)+1)(1-f(E',\varphi))f(E,\varphi)\right],$$
$$v_E^{(R)} = \frac{v_r}{\hbar v_F}E(1-\delta v(\varphi)/v_F), \quad E' = -E + 2E\delta v(\varphi)/v_F \quad , \quad (14)$$

and for negative energies, corresponding to the valence band, $J_R$ can be written in an analogous way. Here $N_{ph}(E,\varphi) = \{\exp[2E(1-\delta v(\varphi)/v_F)/T]-1\}^{-1}$ is the Plank distribution function, $T$ is temperature in energy units, and $v_r = 8\alpha n v_F (v_F/c)^2/3$ is the characteristic radiative velocity.

As we mentioned above, to analyze the recombination current, we need to determine the isotropic component of the distribution function, $f_0(E)$. Restricting ourselves by the linear in strain magnitude contributions to the valley current, we should address this problem assuming no presence of strain. Even in this case this is a complicated problem, requiring, in general, extensive numerical simulations. Below we consider two limiting cases, which allow an approximate solution: the cases of intrinsic and heavily doped graphene.

### 3.2.1. *Intrinsic graphene.*

At low temperatures the concentration of carriers of the intrinsic graphene is small and as a result one can neglect by the carrier-carrier interaction. This case was thoroughly analyzed in [57]. The distribution function at low pumping is split as $f_0(E) = f^{(eq)}(E) + \text{sgn}(E)\Delta f(|E|)$ where $f^{(eq)}(E) = [\exp(E/T)+1]^{-1}$ is the equilibrium distribution and the small nonequillibrium correction $\Delta f(E)$ is determined by the interplay between the thermal radiation generation-recombination processes and the quasielastic energy relaxation due to the acoustic phonon scattering.

After some algebra for the first order in strain contribution to the recombination scattering integral we obtain

$$J_R\{f_0\} = \frac{v_r}{\hbar v_F^2} \delta v(\varphi) E \frac{d}{dE}\left[\frac{E\Delta f(E)}{\sinh(E/T)}\right] \tag{15},$$

which provides the following expression for the recombination valley current

$$\begin{Bmatrix} j_x^{(R)} \\ j_y^{(R)} \end{Bmatrix} = \frac{e}{\pi^2 \hbar^3} \frac{v_r a_2}{v_F^2} \begin{Bmatrix} -\varepsilon_\Delta \\ 2\varepsilon_{xy} \end{Bmatrix} \int_0^\infty dE \frac{E\Delta f(E)}{\sinh(E/T)} \frac{d}{dE}\left[E^2 \tau_1(E)\right]. \tag{16}$$

Naturally, if carrier relaxation due to acoustic phonon scattering is weak, $\Delta f$ is concentrated near $E = E_\omega$ and the absolute value of the recombination current is of the same order as generation one, leading to its partial compensation. However, this is typically not the case [57], and $\Delta f(E)$ is localized in the region $E \sim T$. To analyze importance of the recombination current we have to take into account that particle conservation under the generation-recombination process requires that $\int_0^\infty dE E^2 \Delta f(E) \sinh^{-1}(E/T) = const$ [57]. For elastic scattering by the unscreened Coulomb impurities $\tau_1 \sim E$. Therefore, presence of the extra energy power in the expression for the recombination valley current with respect to the normalization integral suggests that it is considerably less than the generation one. For example, for the same parameters used under calculation of the generation current and $T = 50K$ we obtain the recombination valley current is directed opposite to the generation one, and its absolute value, $j^{(R)} \sim 10^{-5}$ pA/μm, is about two orders of magnitude smaller than $j^{(G)}$. Here we take the same parameters characterizing acoustic phonon scattering, as that used in [57,58], namely, the deformation potential constant $D = 12\text{eV}$, density $\rho_S = 7.6 \cdot 10^{-7} \text{kg/m}^2$ and sound velocity $s = 7.3 \cdot 10^3 \text{m/s}$.

### 3.2.2. *Doped graphene*.

Another case which allows explicit estimate of the recombination valley current is the case of doped graphene. For high enough carrier concentration electron-electron interaction is more efficient than phonon scattering. On the other hand, electron-electron scattering can still be less efficient than elastic scattering by impurities. So, for the Fermi energy $E_F = 34\text{meV}$ which corresponds at low temperatures to the electron concentration $n_e = 10^{15} \text{m}^{-2}$, the charged impurity scattering time can be estimated as $\tau_2 = 5 \cdot 10^{-14} \text{s}$ [54]. For electrons at the chosen excitation energy the electron-electron scattering time is of the order of $0.3 \div 0.1\,\text{ps}$ and is much shorter then the acoustic-phonon scattering times of the order of $5 \div 1\text{ps}$ [59]. For this relaxation times hierarchy it is reasonable to assume that $f_0(E)$ is close to Fermi distribution function but with temperature $T_e$ higher then the lattice and photon temperature $T$ due to the light-induced heating of the electron gas. Naturally, at equilibrium with $\Delta T = T_e - T = 0$ the recombination valley current is zero, and for low excitation power we expect it to be proportional to the ratio $\Delta T / T$. It should be determined from the energy balance which equate the energy input rate due to optical excitation and energy relaxation rate, which in our case is due to acoustic phonon scattering. Note, that in the following we disregard by the light-induced variation of $E_F$ since for weak excitation power it provides no contribution to the valley current. As in the previous subsection, the distribution function splits as $f_0(E) = f^{(eq)}(E) + \Delta f(E)$ where

$f^{(eq)}(E) = [\exp((E-E_F)/T)+1]^{-1}$ is the equilibrium distribution and the small nonequillibrium correction $\Delta f(E) = \dfrac{\partial f^{(eq)}}{\partial T}\Delta T$. For high enough doping with $E_F/T \gg 1$ we obtain the following expressions for the recombination-induced valley current components:

$$\begin{Bmatrix} j_x^{(R)} \\ j_y^{(R)} \end{Bmatrix} \cong \frac{216}{\pi}\frac{ev_r a_2 T^4}{\hbar^3 v_F^2}\begin{Bmatrix} -\varepsilon_\Delta \\ 2\varepsilon_{xy} \end{Bmatrix}\frac{\tau_{1\omega}}{E_\omega}\frac{\Delta T}{T}e^{-E_F/T} \qquad (17)$$

As we mentioned above, the light-induced heating $\Delta T$ is determined from the energy balance equation:

$$\int E^2 [J_{LA}\{f_0(E)\} + J_R\{f_0(E)\} + G]dEd\varphi = 0 \qquad (18)$$

Using the explicit form of the collision integral $J_{LA}$ [57]

$$J_{LA}\{f_0(E)\} = \left(\frac{s}{v_F}\right)^2 \frac{v_{ac}}{\hbar v_F T}\frac{1}{E}\frac{d}{dE}\left[E^4 \frac{df^{(eq)}(E)}{dE}\right]\Delta T \qquad (19)$$

we arrive to the following expression for the light-induced heating:

$$\frac{\Delta T}{T} = \frac{\alpha \pi^2 (\hbar v_F)^3 I_0 t^2}{2 E_F^4 v_{ac}}\left(\frac{v_F}{s}\right)^2 \qquad (20)$$

where $v_{ac} = D^2 T/(4\hbar^2 \rho_S v_F s^2)$ is characteristic acoustic-phonon scattering velocity. The contribution of recombination to the energy balance is negligibly small and the corresponding term is omitted in Eq. (20). Note, that in some actual setups lattice and photon temperatures could be different and this changes the energy balance conditions [60]. Finally, using Eqs.(17,20) we obtain an expression for the ratio of the recombination and generation-induced valley currents:

$$j^{(R)}/j^{(G)} \approx \frac{36}{5} \frac{v_r}{v_{ac}} \left(\frac{v_F}{s}\right)^2 \left(\frac{T}{E_F}\right)^4 e^{-E_F/T} \qquad (21)$$

For chosen above parameters we have $v_{ac} = 1424$ m/s, $v_r = 0.34$ m/s, $\Delta T/T \sim 0.02$ and the current ratio is $j^{(R)}/j^{(G)} \sim 10^{-6}$. So one can make the conclusion that the recombination-induced valley currents in doped graphene are negligibly small compared to the generation ones.

## CONCLUSIONS

To conclude, we analyzed appearance of the valley current in strained graphene under monochromatic optical excitation. The valley current is possible due to the strain-induced electron-hole spectrum anisotropy. Under mid-infrared and softer irradiation for realistic strain magnitudes the considered mechanism of the valley current generation is considerably more efficient than that related to the natural graphene spectrum warping, proposed previously. It is shown that the reverse process of carrier recombination is inessential for valley current formation for both intrinsic and doped graphene due to efficient carrier energy relaxation. The feasible valley current magnitude is about $10^{-3}$ pA/$\mu$m and potentially it can be governed in strain-controlled structures.

## ACKNOWLEDGMENTS

The author thanks V. A. Kochelap for stimulating discussions and acknowledge the hospitality of the Abdus Salam International Centre for Theoretical Physics (Trieste), where this work was completed. This work was supported by the STCU Project №5716.

FIGURE CAPTIONS

**Figure 1.** (a) The honeycomb lattice of graphene. The carbon sites belonging to the two equivalent sublattices are denoted by solid and hollow circles. The dash line marks the primitive cell. (b) The first Brillouin zone of graphene; the Dirac points of the graphene spectrum are at K and K' valleys.

**Figure 2.** Comparison of the strained (solid) with $\varepsilon_{xy} \neq 0$ and unstrained (dot) cross-section of the graphene energy spectrum at $k_x = 0$. The light-induced transitions are marked by vertical arrows, with thickness reflecting the magnitude to the transition probability. The solid arrows of different length indicate the anisotropy of the group velocity.

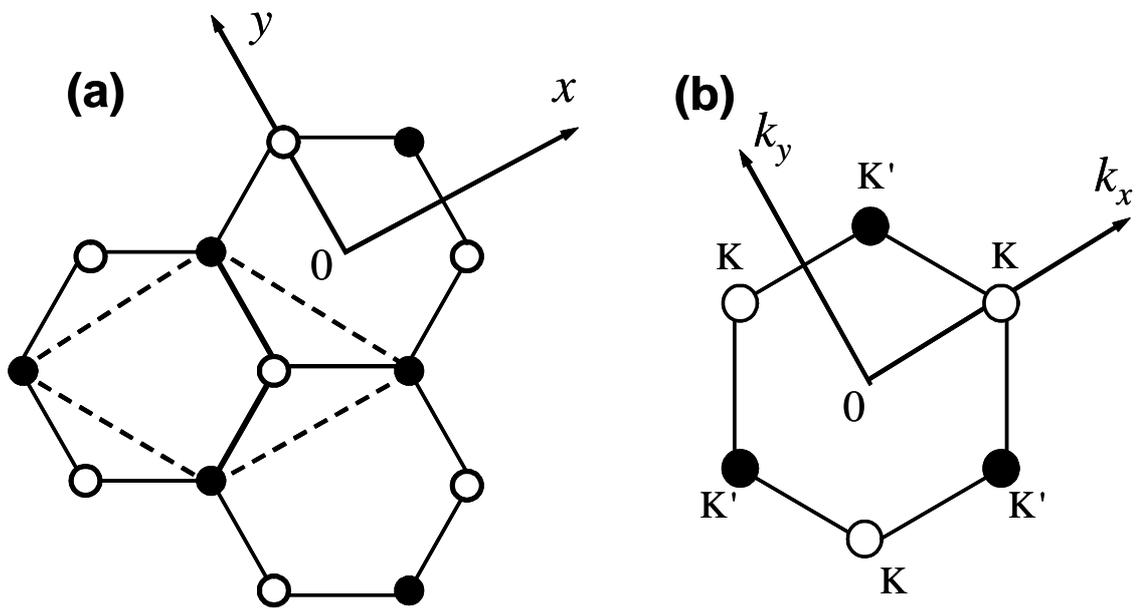

**Figure 1.** T. L. Linnik "Photo-induced valley currents in strained graphene".

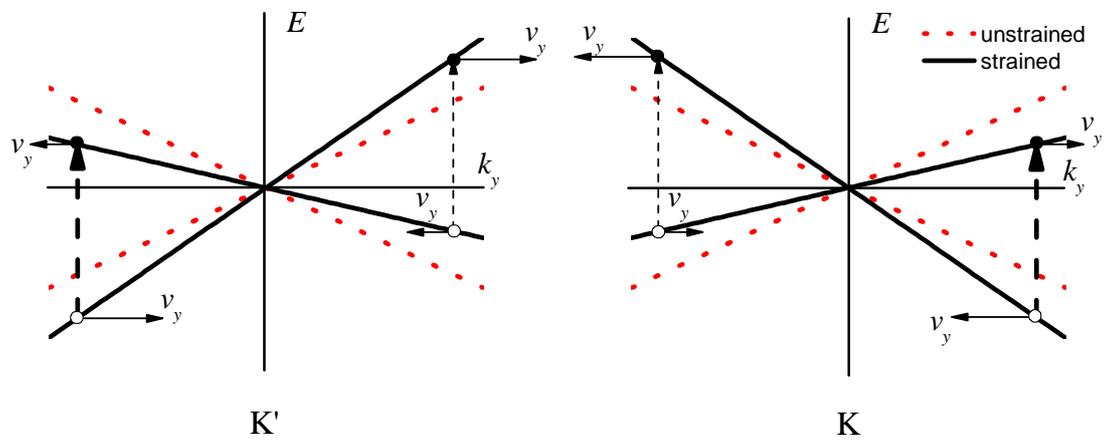

**Figure 2.** *T. L. Linnik* "*Photo-induced valley currents in strained graphene*".